\documentclass[pra,aps,twocolumn,superscriptaddress,longbibliography]{revtex4-1}
\usepackage{amsfonts,amsmath,bm}
\usepackage{color}
\newcommand{\rl}{\rangle\!\langle}
\newcommand{\rr}{\bm{r}}
\newcommand{\RR}{\bm{R}}

\newcommand{\kp}{\mbox{$\bm{k}\!\cdot\!\bm{p}$} }

\usepackage{graphicx}

\DeclareMathOperator{\tr}{Tr}

\usepackage{color}

\usepackage{ulem} % Crossing out with \sout{...}

\begin{document}

\title{Hyperfine interaction for holes in quantum dots: \kp model} 
\author{Pawe{\l} Machnikowski}
\author{Krzysztof Gawarecki}
\affiliation{Department of Theoretical of Physics, Faculty of Fundamental Problems of
  Technology, Wroc{\l}aw University of Science and 
  Technology, 50-370 Wroc{\l}aw, Poland}
\author{{\L}ukasz Cywi\'nski}
\affiliation{Institute of Physics, Polish Academy of Sciences, 02-668 Warsaw, Poland}

\begin{abstract}
We formulate the multi-band \kp theory of hyperfine interactions for semiconductor
nanostructures in the envelope function approximation. We apply this theoretical
description to the fluctuations of the longitudinal and transverse Overhauser field
experienced by a hole for a
range of InGaAs quantum dots of various compositions and geometries. We find that for
  a wide range of values of $d$-shell admixture to atomic states forming the top of the
  valence band,  the transverse Overhauser field caused by this admixture 
is of the same order of magnitude as the longitudinal one, and band mixing adds only a
minor correction to this result.  In consequence, the \kp results are well reproduced by a simple box model with
the effective number of ions determined by the wave function participation number, as long
as the hole is confined in the compositionally uniform volume of the dot, which holds in
a wide range of parameters, excluding very flat dots. 
\end{abstract}

\maketitle

\section{Introduction}
\label{sec:intro}
Hyperfine coupling between the spin of a hole localized in a self-assembled quantum dot
(QD) and
the nuclear spins of the atoms of the host materials 
has been a subject of intense experimental 
\cite{Gerardot_Nature08,Brunner_Science09,Eble2009a,Fallahi_PRL10,Chekhovich_PRL11,%
Chekhovich_NP12,Carter_PRB14,Prechtel_NM16}
and theoretical \cite{Fischer_PRB08,Testelin2009a,Chekhovich_NP12} investigations in
recent years. The original reason for resurgence of interest in this topic was the prospect of
using hole spins as qubits with long coherence times
\cite{Gerardot_Nature08,Brunner_Science09,DeGreve_NP11,Greilich_NP11,Delteil_NP16,%
Prechtel_NM16,Huthmacher_PRB18}.
This was motivated by the fact that dephasing of electron spins in QDs, being an
obstacle to their application as qubits for quantum information processing purposes, is
dominated by their hyperfine (hf) interaction with the nuclear spins of the host material
\cite{Schliemann_JPC03,Coish_PSSB09,Cywinski_APPA11,Urbaszek_RMP13,%
Chekhovich_NM13,Glazov-book}, and 
the hole-nucleus coupling was expected to be much weaker than the electron-nucleus one
\cite{Fischer_PRB08,Testelin2009a}. Experimental confirmation of this expectation
  \cite{Gerardot_Nature08,Brunner_Science09,Eble2009a,Fallahi_PRL10,Chekhovich_PRL11}
  opened the way for using hole spin qubits in applications, such as creation of 
long-distance entanglement of hole spins \cite{Delteil_NP16},  in which their enhanced
coherence time and good coupling to photons (holding for both holes and electrons in
self-assembled QDs \cite{DeGreve_RPP13,Warburton_NM13}) was helpful. 

The dominating mechanism in the case of electron hf coupling is the contact
interaction \cite{dyakonov72,Schliemann_JPC03}. 
While for a purely $s$-shell state this interaction would be isotropic, 
lowering the local symmetry leads to mixing of atomic shells in the Bloch functions,
which results in an anisotropy of the hf coupling due to dipolar coupling to non-$s$
atomic states
\cite{Gryncharova1977,Obata1963,Hale_PRB71}, as in the case
of electron states in materials such as Si, in which the states at the bottom of
conduction bands have appreciable non-$s$ component \cite{Hale_PRB71}.
On the contrary, the holes have very small
contribution of $s$ states in their wave functions, and therefore interact with nuclei 
only by
weaker (approximately by one order of magnitude \cite{Fischer_PRB08}) and 
much more subtle dipole couplings, which are sensitive to the details of their
  atomic (or Bloch) wave functions.
 For non-$s$ states, anisotropy can be induced by symmetry breaking
  without shell mixing, by modifying the hybridization of orbitals. In particular, breaking
  the symmetry on the mesoscopic level by strong confinement in the growth ($z$-axis)
  direction,  lifts the degeneracy between heavy hole and light hole states
 and leads to strong anisotropy of the hole hf interaction. As a
  consequence, for purely heavy-hole state, and for Bloch functions at the top of the
valence band being 
built only from atomic $p$-shell orbitals of the atoms constituting the crystal, the hf
interaction should be of Ising character, with the interaction axis parallel to the growth
axis \cite{Fischer_PRB08}. However, similar to the electron case, atomic shell
  mixing, in particular the finite amplitude of $d$ states at the top of the valence band
(i.e.~the $p$-$d$ hybridization) can give rise to transverse couplings. Such
 couplings appear also in the presence of finite 
heavy-light hole mixing \cite{Fischer_PRB08,Eble2009a,Testelin2009a}. 

These general expectations were confirmed by experiments, showing that the Overhauser field
exerted by the nuclei on the hole is about an order of magnitude smaller than the one
experienced by the electron in the same dot. 
In \cite{Gerardot_Nature08} and \cite{Brunner_Science09}
  qualitative results showing  that in InGaAs QDs the coupling of holes to the nuclei is
  much weaker than that of electrons were obtained from analysis of hole spin
  initialization by optical pumping, and 
  coherent population trapping experiment, respectively. In \cite{Eble2009a}
  photo-induced 
  circular dichroism of an ensemble of QDs was measured and, from its magnetic field
  dependence, the value of dot ensemble-average of {\it transverse} Overhauser field
  experienced by the hole spin was estimated to be $\approx\! 30$ times smaller than the
  field experienced by the electron spin, and theoretical estimates suggested that the
  longitudinal coupling should be larger by a factor of about $2$, implying rather weak
  anisotropy of the interaction. Direct measurements of relative magnitudes of the Overhauser
  field experienced by electrons and holes were then described in \cite{Fallahi_PRL10} and
  \cite{Chekhovich_PRL11}, where nuclei in single InGaAs/GaAs and InP/GaInP single QDs,
  respectively, were dynamically polarized, and the resulting splittings of electron and hole
  spin states were measured. In both experiments the magnitude of the {\it longitudinal} hole
  Overhauser field was $\approx\! -0.1$ of the electron field, in qualitative agreement
  with theoretical predictions \cite{Fischer_PRB08}. It is important to note, however,
  that the anisotropy of the hole hf interaction was quantitatively characterized in experiments
  concerning the same quantum dot only very recently in \cite{Prechtel_NM16}, where a value
  of $\sim \! 1$ \% of the longitudinal one was measured, while the latter was about
  $10$\% of the electron hf interaction, in agreement with previous experiments. 

Clearly, the existing experiments do not paint a fully consistent picture, and a
number of open questions and controversies needs to be investigated. While the
longitudinal (along 
the QD growth axis) hole-nuclear interaction qualitatively agreed with theoretical
expectations, large discrepancies in strength of transverse hyperfine interactions were 
reported, with estimated values of transverse coupling ranging from the same order of
magnitude \cite{Eble2009a} to less than $1\%$ of the longitudinal one
\cite{Prechtel_NM16}. 
The origin of the transverse interactions is also controversial: while
initially heavy-light hole mixing was invoked \cite{Fischer_PRB08,Eble2009a,Testelin2009a}
to explain its finite value,  the presence of finite admixture of $d$-symmetry states in
states forming the top of valence band in relevant III-V materials was suggested to
play a significant, or even possibly dominant, role \cite{Chekhovich_NP12}. Such a
substantial admixture of atomic $d$ states in the valence band Bloch functions is in
qualitative accordance with earlier theoretical results
\cite{Diaz2006,Chadi1976,Persson2003,Boguslawski_SST94}. However, as we show in this
  paper, the admixture of $d$ states used in  \cite{Chekhovich_NP12} to explain the
  relative magnitudes and signs of contributions to the {\it longitudinal} Overhauser
  field coming from various atoms, implies that the equilibrium fluctuations of the {\it
    transverse} Overhauser field should be comparable to those of the longitudinal one,
  leading to an apparent qualitative contradiction with the results of
  \cite{Prechtel_NM16}. 
 
It is important to note that with transverse hole hyperfine coupling being much smaller
than the longitudinal one (which, in turn, is lower by an order of magnitude than the
coupling for an electron), the coherence time of hole spin polarized along the growth axis
can be significantly enhanced by application of large transverse magnetic field
perpendicular to this axis
\cite{Fischer_PRB08,Prechtel_NM16}. Understanding of physical origin of the transverse
coupling is thus important, as it would possibly allow for design of QDs (by
varying composition/shape/strain etc.) with the best possible hole spin coherence
properties.   

The motivation for this work is the observation that interpretation of most
experiments related to physics of 
carrier and nuclear spins in QDs  relies on simplified models of carrier envelope
wave functions (e.g.~assuming the same envelope shapes for holes and electrons) and all the
multi-band effects (including the degree of heavy-light hole mixing), and their relation
to QD shape. This includes works on carrier spin coherence (which apart from
nuclear effects show influence of charge noise coupling to spin via electric-field
dependent $g$-factors \cite{DeGreve_NP11,Huthmacher_PRB18}), creation of dynamic nuclear
polarization \cite{Fallahi_PRL10,Chekhovich_PRL11}, and optical detection (through changes
in Overhauser field-induced spin splitting of electron and holes) of nuclear magnetic
resonance of different species of nuclei present in the dot \cite{Chekhovich_NN12}. 
While such experiments were used to obtain new information on structural properties and
strain distribution in QDs \cite{Chekhovich_NN12}, the simplicity of some of the
above-mentioned assumptions casts a certain degree of doubt on the interpretation of
measurement results. In light of the above-discussed disagreements between distinct experiments, more careful studies of hole states and the hyperfine coupling for holes are clearly necessary.

The current state of the art in the theoretical modeling of self-assembled semiconductor
structures is to use either atomistic methods \cite{bester05,Wei2014,Zielinski2012}, or
multi-band \kp 
theories in the envelope function approximation  \cite{ehrhardt14,Gawarecki2018}. The latter has found
a vast range of applications due to its relatively low computational cost and high
versatility. It offers reliable information on the
wave function geometry and band mixing and allows one to quantitatively relate the
observed spectral features to fine details of the nano-system morphology and composition.
It can be used not only to compute the carrier states and the resulting
optical transitions \cite{Schliwa2007}, but also to model carrier-phonon couplings
\cite{Gawarecki2012} and to evaluate the spin-related properties, including $g$-factors
\cite{jovanov12,Andlauer2009}, 
the effects of spin-orbit coupling  \cite{winkler03,Gawarecki2018}, as well as phonon-induced spin
relaxation and dephasing \cite{Mielnik-Pyszczorski2018a,Gaweczyk2018}. Therefore, in
terms of quantitative accuracy, a simple approach to hyperfine interactions lags behind
the current standards 
in the modeling of carrier wave functions in semiconductor nanostructures and  is not on 
a par with the sophistication of experimental techniques used for the measurements of
the relevant quantities. 
It therefore seems useful to develop a theory that
would allow one to combine the hyperfine interaction with realistic modeling of wave
functions.  Such a more general and accurate theory may be useful in systems with 
compositional inhomogeneity and controllable carrier localization, like double QDs, or 
with strong in-plane anisotropy, where band mixing is relatively stronger
\cite{Musial2012}.

The goal of this paper is to revisit the problem of calculation of the anisotropic
Overhauser field acting on a hole spin while employing a detailed realistic description of
carrier states in QDs.  
We derive a theoretical description of hyperfine coupling for 
a carrier confined in a self-assembled semiconductor QD
 based on the multi-band wave function
obtained from the \kp theory in the envelope function approximation, taking into account
d-wave admixture in the valence-band states. In this way, we provide a model of the
hyperfine interaction compatible with the standard \kp modeling of carrier states, which
opens the way towards combining the effects of hyperfine coupling with reliable modeling
of other characteristics of the QD system.
As an application of the formalism, we
calculate the rms fluctuations of the 
longitudinal and transverse Overhauser field in InGaAs/GaAs QDs and compare the
contributions to the 
transverse field fluctuations from band
mixing and d-wave admixture to valence band states. 

The paper is organized as follows. In Sec.~\ref{sec:hamiltonian} we derive the general
8-band \kp Hamiltonian for hyperfine interactions. Next, in Sec.~\ref{sec:hf-hole} we
apply this formalism to the fluctuations of the Overhauser field felt by a hole in a QD. In Sec.~\ref{sec:discussion} we discuss the implications that our results have on hole spin decoherence and the status of experimental controversies concerning hf interaction of holes, and
in Sec.~\ref{concl} we summarize our findings. Technical derivations are collected in the Appendix.

\section{The multi-band hyperfine Hamiltonian}
\label{sec:hamiltonian}

The hyperfine Hamiltonian describes the interaction of the carrier with all the nuclei
(labeled by $\alpha$ and located at $\RR_{\alpha}$),
\begin{equation}\label{Hhf}
H=3E_{\mathrm{hf}}
\sum_{\alpha} \zeta_{\alpha}
\bm{A}(\rr-\RR_{\alpha})\cdot\bm{I}_{\alpha}/\hbar,
\end{equation}
where 
\begin{displaymath}
E_{\mathrm{hf}} =
\frac{2\mu_{0}}{3\pi}\mu_{\mathrm{B}}\mu_{\mathrm{N}}
a_{\mathrm{B}}^{-3} = 0.5253\,\mu\mathrm{eV},
\end{displaymath}
$\mu_{\mathrm{B}}$ and $\mu_{\mathrm{N}}$ are Bohr and nuclear magnetons, respectively,
$a_{\mathrm{B}}$ is the Bohr radius, $\mu_{0}$ is the vacuum permeability,
$\bm{I}_{\alpha}$ is the nuclear spin, $\zeta_{\alpha}$ defines the nuclear
magnetic moment for a given nucleus via
$\bm{\mu}_{\alpha}=\zeta_{\alpha}\mu_{\mathrm{N}}\bm{I}_{\alpha}$, and   
\begin{equation}\label{A-all}
\bm{A}(\rr) = \frac{a_{\mathrm{B}}^{3}}{4\hbar}\left[
\frac{8\pi}{3}\delta(\rr)\bm{S} + \frac{\bm{L}}{r^{3}}
+\frac{3(\hat{\rr}\cdot\bm{S})\hat{\rr}-\bm{S}}{r^{3}}  \right],
\end{equation}
with $\bm{L}$ and $\bm{S}$ denoting the orbital and spin angular momentum of the carrier
and $\hat{\rr}=\rr/r$. The first term in Eq.~\eqref{A-all} is the Fermi contact
interaction between the carrier and nuclear spins, the second term describes the coupling
of the nuclear spin
to the electric current associated with the orbital motion of the carrier, and the last one
is the dipole interaction between the nuclear and carrier spins.

Within the envelope function approach to the
\kp theory, the wave functions are decomposed into
contributions from various bands 
$\lambda$ with $\Gamma$-point Bloch functions $u_{\lambda}(\rr,s)$,
\begin{equation}\label{Psi}
\Psi_{\nu}(\rr,s)=\sum_{\lambda}\psi_{\nu,\lambda}(\rr) u_{\lambda}(\rr,s),
\end{equation}
where the envelopes $\psi_{\nu,\lambda}(\rr)$ are assumed to vary slowly in space (as
compared to the lattice constant) and
$s$ denotes the spin projection. Most commonly, an 8-band model is used \cite{winkler03},
explicitly representing two subbands of the conduction band (belonging to the 
$\Gamma_{6\mathrm{c}}$ representation of the bulk crystal) and six subbands in the valence
band (four-dimensional $\Gamma_{8\mathrm{v}}$ and two-dimensional $\Gamma_{7\mathrm{v}}$),
with the coupling to other bands represented by effective terms resulting from 
perturbation theory.  The eight envelope wave functions $\{\psi_{\nu,\lambda}(\rr)\}$ are
commonly thought of as an
8-component ``pseudo-spinor''. Consequently, the Hamiltonian (or any other operator) in
the envelope-function \kp theory can be considered an $8\times 8$ array of operators
$H_{\lambda'\lambda}$ in the coordinate representation, such that any matrix element of
the original Hamiltonian is given by  
\begin{equation}\label{matelem-kp}
\langle\nu | H | \mu \rangle = \sum_{\lambda'\lambda} 
\int d^{3}r \, \psi^{*}_{\nu,\lambda'}(\rr) H_{\lambda'\lambda} \psi_{\mu,\lambda}(\rr).
\end{equation}
The goal of this section is to apply the envelope function approximation [Eq.~\eqref{Psi}] to the hyperfine
Hamiltonian \eqref{Hhf} and to write it in the form consistent with
Eq.~\eqref{matelem-kp}. 

Starting from Eq.~\eqref{A-all} and using Eq.~\eqref{Psi}, the matrix elements of $A_{i}$
are  
\begin{align}\label{m-elem-A}
\langle \nu | A_{i}(\rr-\RR_{\alpha}) | \mu\rangle &= 
\sum_{\lambda'\lambda} \sum_{ss'} \int d^{3}r \,
\psi^{*}_{\nu,\lambda'}(\rr) u^{*}_{\lambda'}(\rr,s') \\
&\quad\times A_{i,s's}(\rr-\RR_{\alpha})
\psi_{\mu,\lambda}(\rr) u_{\lambda}(\rr,s), \nonumber
\end{align}
where  $A_{i,s's}(\rr)$ denotes the matrix elements of $A_{i}(\rr)$ with
respect to spin states.
The Bloch
functions are decomposed into parts localized around the anion (A) and
cation (C), that are assumed to be normalized and non-overlapping,
\begin{displaymath}
u_{\lambda}(\rr,s)=\sum_{i=\mathrm{A,C}}a_{i}^{(\lambda)} u_{\lambda}^{(i)}(\rr,s),
\end{displaymath}
where
$a_{\mathrm{A,C}}^{(\lambda)}$ are the contributions of the anionic and
cationic atomic orbitals to a given band.
Next, we split the space into primitive cells, which are further divided into two parts
surrounding the anion and the cation. The integration over the whole space  is then
performed as integration over the surrounding of each ion and summation over all the
ions. We use the fact that the envelope
varies slowly, so that in the vicinity of each ion it can be approximated by its value
at the ion position $\bm{R}$. In this way we transfer
Eq.~\eqref{m-elem-A} into
\begin{equation}\label{m-elem-A-sum}
\langle \nu | A_{i}(\rr-\RR_{\alpha}) | \mu\rangle =
v \sum_{\lambda'\lambda}\sum_{\alpha'} 
\psi^{*}_{\nu,\lambda'}(\bm{R}_{\alpha'}) A_{i,\lambda'\lambda}^{\alpha'\alpha}
\psi_{\mu,\lambda}(\bm{R}_{\alpha'}), 
\end{equation}
with
\begin{equation}\label{All}
A_{i,\lambda'\lambda}^{\alpha'\alpha} = \frac{1}{v}
\sum_{ss'}\int_{V_{\alpha'}} d^{3}r \, u^{*}_{\lambda'}(\rr,s') A_{i,s's}(\rr-\RR_{\alpha})
 u_{\lambda}(\rr,s),
\end{equation}
where $v$  is the volume of the primitive crystal cell and
$V_{\alpha}$ denotes the volume surrounding the ion $\alpha$ (the arbitrariness in
choosing this volume is unimportant in view of the strong localization of Bloch
functions around the ions \cite{Chekhovich_NP12}). Since the variation of the envelope
functions is slow, the summation in Eq.~\eqref{m-elem-A-sum} realizes a coarse-grained integration
over the whole space. Thus, Eq.~\eqref{m-elem-A-sum} brings matrix elements of the
hyperfine Hamiltonian, Eq.~\eqref{Hhf}, to the form of Eq.~\eqref{matelem-kp} with
\begin{equation}\label{H-kp}
H_{\lambda'\lambda}(\rr) = 3E_{\mathrm{hf}}
v\sum_{\alpha'\alpha} \delta(\rr-\RR_{\alpha}) \zeta_{\alpha}
\bm{A}^{\alpha'\alpha}_{\lambda'\lambda}\cdot\bm{I}_{\alpha}/\hbar,
\end{equation}

In order to evaluate Eq.~\eqref{All} one needs a model of the Bloch functions. Following
\cite{Fischer_PRB08}, we choose to represent them as combinations of
normalized hydrogen-like functions $f_{lm}^{(i)}(\rr)$
with definite rotational symmetry ($l=s,p,d$), characterized by 
the orbital
exponents $\xi_{l,\alpha}$  \cite{Clementi1963,Clementi1967} that depend on the nuclear
species occupying the site $\alpha$. 
Thus, 
\begin{equation}\label{Bloch-i}
u_{\lambda}^{(i)}(\rr,s)=\sqrt{v}
\sum_{lm}c^{(\lambda s)}_{lm}f_{lm}^{(i)}(\rr-\rr_{i}),
\end{equation}
where $l=0,1,2,\;m=-l,\ldots,l$. 
The valence band Bloch functions are composed of $p$ and $d$ atomic orbitals, weighted by
the amplitudes $\alpha_{p}$ and $\alpha_{d}$, respectively, with
$|\alpha_{p}|^{2}+|\alpha_{d}|^{2}=1$. 
We suppress the principal quantum number $n$ since only one orbital of each symmetry is
relevant for a given atom.
The coefficients 
$c^{(\lambda s)}_{lm}$ for purely $p$-band ($l=1$) states can be found from
angular momentum addition and 
are widely available in the literature related to the \kp method
\cite{winkler03,Willatzen2009}. The 
extension to the $d$ admixture follows immediately from the explicit form of the basis
functions of the $F_{2}$ representation of the $T_{d}$ point group, as
given in \cite{Chekhovich_NP12}. 

The matrix element in Eq.~\eqref{H-kp} has
two contributions: the local, or short-range (SR) one, from the
surrounding of the ion in 
question ($\alpha'=\alpha$) and the long-range (LR) one, from all
the other ions in the crystal (including the neighboring cations for an anion and
vice-versa). The LR contribution has been estimated to be negligible  
\cite{Obata1963,Hale_PRB71,Fischer_PRB08,Testelin2009a}. In the following we only take into account the SR
contribution.

The detailed derivation of the SR contributions, which systematically
extends the existing theoretical description
\cite{Gryncharova1977,Fischer_PRB08,Eble2009a,Chekhovich_NP12} to multi-band wave functions,
is given in the Appendix.  
The resulting matrix elements $ A_{i,\lambda'\lambda}^{\alpha\alpha}$ must have
appropriate transformation properties, hence they can be expressed by the standard
matrices used to define point group invariants when constructing the \kp theory. In order
to use this convenient notation we split
the array $\{H_{\lambda'\lambda}\}$ into blocks corresponding to the three irreducible
representations spanning the 8-band \kp model,
\begin{equation}\label{Ham-kp}
H = \left(\begin{array}{ccc}
H_{\mathrm{6c6c}} & H_{\mathrm{6c8v}} & H_{\mathrm{6c7v}} \\
H_{\mathrm{8v6c}} & H_{\mathrm{8v8v}} & H_{\mathrm{8v7v}} \\
H_{\mathrm{7v6c}} & H_{\mathrm{7v8v}} & H_{\mathrm{7v7v}} 
\end{array}\right),
\end{equation}
with 
\begin{equation} \label{Hbb}
H_{b'b} = H^{\dag}_{b'b}
=E_{\mathrm{hf}} v\sum_{\alpha} \delta(\rr-\RR_{\alpha})  a_{\alpha}^{(b')*}a_{\alpha}^{(b)}
\zeta_{\alpha}  \xi_{s,\alpha}^{3} \tilde{H}_{b'b}^{(\alpha)}
\end{equation}
(the index $b$ labels blocks and we assume that $a_{\alpha}^{(\lambda)}$ is the same for
all bands $\lambda$ in a given block $b$), 
and find 
\begin{subequations}
\begin{align}
\label{H6c6c}
\tilde{H}_{\mathrm{6c6c}}^{(\alpha)} & = \bm{\sigma}\cdot \bm{I}_{\alpha}/\hbar, \\
\label{H8v8v}
\tilde{H}_{\mathrm{8v8v}}^{(\alpha)} & = \left(-\frac{8}{5}\tilde{M}_{p}^{(\alpha)} +
  \frac{39}{7} \tilde{M}_{d}^{(\alpha)} \right) \bm{J}\cdot \bm{I}_{\alpha}/\hbar \\
&\quad -\frac{12}{7} \tilde{M}_{d}^{(\alpha)}
\bm{\mathcal{J}}\cdot \bm{I}_{\alpha}/\hbar, \nonumber \\
\tilde{H}_{\mathrm{7v7v}}^{(\alpha)} & = \left(-4 \tilde{M}_{p}^{(\alpha)}
+\frac{2}{7}\tilde{M}_{d}^{(\alpha)} \right)
 \bm{\sigma}\cdot \bm{I}_{\alpha}/\hbar, \\
\tilde{H}_{\mathrm{6c8v}}^{(\alpha)} & = -\frac{9}{\sqrt{5}}\tilde{M}_{sd}^{(\alpha)} 
\left(T_{xy}I_{\alpha,z} + T_{yz}I_{\alpha,x} + T_{zx}I_{\alpha,y} \right)/\hbar, \\
\tilde{H}_{\mathrm{6c7v}}^{(\alpha)} & = 0, \\
\label{H7v8v}
\tilde{H}_{\mathrm{7v8v}}^{(\alpha)} & = -\sqrt{3}
\left(\tilde{M}_{p}^{(\alpha)} -\frac{15}{7}\tilde{M}_{d}^{(\alpha)}\right)
\bm{T}\cdot \bm{I}_{\alpha}/\hbar,
\end{align}
\end{subequations}
where $\tilde{M}_{p,d}^{(\alpha)}=|\alpha_{p,d}^{(\alpha)}|^{2}M_{p,d}^{(\alpha)}$, 
$\tilde{M}_{sd}^{(\alpha)}=\alpha_{d}^{(\alpha)} M_{sd}^{(\alpha)}$,
the dimensionless quantities $M_{p,d,sd}$ characterize the geometry of the atomic functions
and are explicitly defined in the Appendix, 
 $\bm{\sigma}=(\sigma_{x},\sigma_{y},\sigma_{z})$ are Pauli matrices,
$\bm{J}=(J_{x},J_{y},J_{z})$ are the matrices of the 4-dimensional ($j=3/2$)
irreducible representation of angular momentum, 
$\bm{\mathcal{J}}=(J_{x}^{3},J_{y}^{3},J_{z}^{3})$,
\begin{align*}
T_{x} & =\frac{1}{3\sqrt{2}}\left(\begin{array}{cccc}
-\sqrt{3} & 0 & 1 & 0 \\ 0 & -1 & 0 & \sqrt{3}
\end{array}\right),\\
T_{y} & =\frac{-i}{3\sqrt{2}}\left(\begin{array}{cccc}
\sqrt{3} & 0 & 1 & 0 \\ 0 & 1 & 0 & \sqrt{3}
\end{array}\right), \quad
T_{z} =\frac{\sqrt{2}}{3}\left(\begin{array}{cccc}
 0 & 1 & 0 & 0 \\ 0 & 0 & 1 & 0 
\end{array}\right), 
\end{align*}
and $T_{ij} = T_{i}J_{j}+T_{j}J_{i}$.
Here the equation for $\tilde{H}_{\mathrm{8v8v}}$ reproduces the result of
\cite{Chekhovich_NP12}. 

From Eq.~\eqref{H8v8v} it is clear that for a purely hh state the only flip-flop terms
appear as a result of $d$-shell admixture via the $J_{i}^{3}$ terms that reflect the
lowered symmetry of the crystal as compared to the full rotation group. As we show in the
Appendix, these terms originate from the spin part of the dipole hyperfine coupling (the
last term in Eq.~\eqref{A-all}). Inter-band terms in the Hamiltonian lead also to
flip-flop processes induced by band mixing but, as we will see below, this effect is much
weaker.

\section{Hyperfine coupling for the heavy-hole ground state} 
\label{sec:hf-hole}

In this section we apply the general formalism of Sec.~\ref{sec:hamiltonian} to the ground
state Zeeman doublet of the nominally heavy hole state in a range of self-assembled QDs
with varying size, shape and composition. We characterize the fluctuations of the
Overhauser field felt by the hole that is the key factor determining the hyperfine-induced
spin dephasing. 

\subsection{QD model and wave functions}
\label{QD}

The envelope functions for the QD ground state  are computed  for a few series of QD
structures with the 8-band \kp
theory. In all the cases the composition of the QD is uniform and corresponds to the
stoichiometric formula $\mathrm{In}_{x}\mathrm{Ga}_{1-x}\mathrm{As}$. The QD is placed on
a wetting layer of the same composition and thickness equal to the GaAs lattice constant
$a=0.565$~nm.  

We account for the strain within continuous-elasticity approach \cite{Pryor98b}. We take into account the
piezoelectric potential, up to the second order in polarization \cite{Bester06a}. 
The magnetic field enters via Peierls substitution within the gauge invariant scheme,
described in detail in \cite{Andlauer08}. 
The detailed description of the model as well as parameters used in computations are given in
\cite{Gawarecki2018}. 

\begin{table}
\caption{\label{params}Nuclear \cite{Schliemann_JPC03} and atomic
parameters.}
\begin{center}
\begin{tabular}{l|ccccc}
 & $^{69}$Ga & $^{71}$Ga & $^{113}$In & $^{115}$In & $^{75}$As \\
\hline
$I$ & $3/2$  & $3/2$ & $9/2$ & $9/2$ & $3/2$ \\
$\zeta$ & 1.344 & 1.708 &  1.227 & 1.230 & 0.959\\
%$I\zeta_{\alpha}$ & 2.016 & 2.562 & 5.523 & 5.534 & 1.439 \\
$r$ & 0.604 & 0.396 & 0.0428 & 0.9572 & 1 \\
\hline
$\xi_{s}$ & \multicolumn{2}{c}{3.9} & \multicolumn{2}{c}{3.9} 
                         & 4.4 \\
$\xi_{p}$ & \multicolumn{2}{c}{3.3} & \multicolumn{2}{c}{3.3} 
                         & 3.7 \\
$\xi_{d}$ & \multicolumn{2}{c}{10.5} &  \multicolumn{2}{c}{8.9} 
                         & 11.9 \\
\hline
$M_{p}$ & \multicolumn{2}{c}{0.050} & \multicolumn{2}{c}{0.050} 
                         & 0.050 \\
$M_{d}$ & \multicolumn{2}{c}{0.33} & \multicolumn{2}{c}{0.20} 
                         & 0.33 \\
$M_{sd}$ & \multicolumn{2}{c}{0.048} &  \multicolumn{2}{c}{0.034} 
                         & 0.049 \\
\hline
$|\alpha_{d}|^{2}$ & \multicolumn{2}{c}{0.20} &  \multicolumn{2}{c}{0.50} 
                         & 0.05 \\
\hline
$|a_{{\mathrm{C/A}}}^{(\mathrm{cb})}|^{2}$ & \multicolumn{4}{c}{0.50} 
                         & 0.50 \\
$|a_{{\mathrm{C/A}}}^{(\mathrm{vb})}|^{2}$ & \multicolumn{4}{c}{0.35} 
                         & 0.65 \\
\hline 
$A^{(\mathrm{e})}\,\mu$eV &
41.9 & 53.2 & 38.2 & 38.3 & 42.9

\end{tabular}
\end{center}
\end{table}

Recently, the exponents of the atomic basis functions
were related to measurable crystal properties \cite{Benchamekh2015} via tight-binding
calculations. However, the Slater orbitals commonly used in the tight-banding models are
inappropriate for calculating the hyperfine effects, as they only capture the asymptotic
behavior of the wave functions away from the nucleus, and are all zero (even those representing the $s$
states) at the position of the nucleus. Thus, although the results of
\cite{Benchamekh2015} show some promise for more accurate modeling of the Bloch
functions, for our purpose we still need to find an appropriate parametrization of the
wave functions. We do so by requiring consistency with the available experimental and
theoretical data: 
the hole-to-electron ratio of Overhauser fields \cite{Chekhovich_NP12}, Ga and As wave
functions at the nucleus \cite{Chekhovich_NM17}, and $d$-shell admixture and anion-cation
distribution in GaAs \cite{Boguslawski_SST94}. With the scarce quantitative data available, the
parametrization remains to a large extent underdetermined. Based on the relations of the Slater
exponents \cite{Clementi1963,Clementi1967,Benchamekh2015}, we arbitrarily set the $s$-shell
exponents for In the same as for Ga and assume  $\xi_{p} = 0.85\xi_{s}$ for all atoms. The
$d$-shell exponents are then determined from the data of
\cite{Chekhovich_NP12}. This parametrization is still to a large extent
arbitrary, and should be considered a starting point for further improvements as new
experimental and computational data become available.

Table~\ref{params} lists the proposed values of the parameters relevant for the modeling
of Bloch functions as well as those describing the hyperfine couplings (see
Sec.~\ref{sec:hamiltonian} and Appendix): nuclear 
spin quantum numbers, $\zeta$ coefficients and relative abundances $r$ for the nuclei
of interest, the atomic wave function exponents $\xi$ and the resulting $M$
parameters, as well as the $d$-state admixture amplitudes $|\alpha_{d}|^{2}$ and
cation-anion distributions of charge density for the conduction and valence bands
($|a_{{\mathrm{C/A}}}^{(\mathrm{cb})}|^{2}$ and $|a_{{\mathrm{C/A}}}^{(\mathrm{vb})}|^{2}$,
respectively). At the bottom of Tab.~\ref{params} we list the resulting values of the
electron hyperfine coupling constant $A^{(\mathrm{e})}=2E_{\mathrm{hf}}|a_{\mathrm{C}}|^{2}\zeta\xi_{s}^{3}$
for each atom, which for Ga and As are very close to those determined in
\cite{Chekhovich_NM17}.

The proposed model is a combination of a standard \kp approach to computing the envelope
wave functions, and a model of atomic wave functions that is necessary for the calculation
of the hyperfine couplings. Although the latter must be done on the atomistic level, the
\kp model itself is not atomistic and remains at the usual mesoscopic level: the strain is
treated within a continuous approach, and the standard values of parameters are used,
unrelated to the model of Bloch functions used in the second stage. In the \kp calculation,
alloying is taken into account in a coarse-grained manner, by interpolating parameters
according to the local composition (virtual crystal approximation), while in the hf
calculations explicit counting of ions 
forces us to implement a particular distribution of atoms and isotopes and to average over
a few realizations of the alloy disorder.

\subsection{Effective Hamiltonian}
\label{sec:effective}

We find the effective Hamiltonian describing the hyperfine interactions in the heavy hole
ground state by projecting Eq.~\eqref{Hhf} onto the two-dimensional space of the
ground-state doublet. We denote the eigenstates in the ground state doublet (as resulting
  from the \kp diagonalization) by
\mbox{$|\!\uparrow\rangle$} and \mbox{$|\!\downarrow\rangle$} (hence the two basis states
are defined with respect to the spin quantization axis) and define operators $\Sigma_{i}$
corresponding to Pauli matrices in
this two-dimensional subspace, 
$\Sigma_{z}=|\!\uparrow\rl\uparrow\!| - |\!\downarrow\rl\downarrow\!|$
etc.
The Hamiltonian given in Eq.~\eqref{Hhf} is linear in the nuclear spins, hence its
projection on the two-dimensional subspace can
be written as 
\begin{equation}\label{H-ef}
H= 
\frac{1}{2}\sum_{\alpha}\sum_{ij}\mathcal{H}_{ij}^{(\alpha)} (I_{\alpha,i}/\hbar)\Sigma_{j},
\end{equation}
where 
\begin{displaymath}
\mathcal{H}_{ij}^{(\alpha)} =3E_{\mathrm{hf}}
\zeta_{\alpha} \xi_{s,\alpha}^{3} \tr \left[ A_{i}(\rr-\RR_{\alpha})\Sigma_{j}\right].
\end{displaymath}
This Hamiltonian has the form of a Zeeman Hamiltonian,
\begin{displaymath}
H = \frac{1}{2}\bm{h}\cdot\bm{\Sigma},
\end{displaymath}
with the quantity $\bm{h}$, defining the Overhauser field, with components given by
\begin{displaymath}
h_{j} = \sum_{\alpha} \sum_{i} \mathcal{H}_{ij}^{(\alpha)}I_{\alpha,i}/\hbar.
\end{displaymath}
We assume here that the nuclei are in a thermal state without any dynamical
polarization. Except for unrealistically low temperatures, this means that the nuclear density
matrix is maximally mixed. 
The mean square of a given component of $\bm{h}$ is then given by 
\begin{align*}
\left\langle h_{j}^{2} \right\rangle
& = \sum_{\alpha\alpha'}\sum_{ii'} \mathcal{H}_{ij}^{(\alpha)}\mathcal{H}_{i'j}^{(\alpha')}
\left\langle I_{\alpha,i} I_{\alpha',i'} \right\rangle/\hbar^{2} \\
&=\frac{1}{3}\sum_{\alpha}I_{\alpha}(I_{\alpha}+1)\sum_{i} \left(\mathcal{H}_{ij}^{(\alpha)} \right)^{2},
\end{align*}
where the last equality assumes that angular momenta of different nuclei as well as
different components of nuclear spin are uncorrelated.

In the simplest approximation, one considers a purely heavy-hole
wave function which occupies a region of uniform composition and is the same for both
  spin orientations. Then, by direct inspection
of Eq.~\eqref{H8v8v} one finds  
\begin{align*}
\mathcal{H}_{ii}^{(\alpha)} & =
2v  E_{\mathrm{hf}} |\psi(\RR_{\alpha})|^{2} \zeta_{\alpha}  \xi_{s,\alpha}^{3}  
\mathcal{M}_{i}^{(\alpha)}, \\
\mathcal{H}_{ij}^{(\alpha)} & = 0,\quad i\neq j,
\end{align*}
 where
\begin{align*}
\mathcal{M}_{x}^{(\alpha)} = \mathcal{M}_{y}^{(\alpha)} & = 
  \frac{9}{7} \tilde{M}_{d}^{(\alpha)} , \\
\mathcal{M}_{z}^{(\alpha)} & =
\frac{12}{5}\tilde{M}_{p}^{(\alpha)} -
  \frac{18}{7} \tilde{M}_{d}^{(\alpha)}.
\end{align*}
Since $M_{i}^{(\alpha)}$ depends only on the species of the ion $\alpha$ and $\psi(\RR)$
changes slowly, one can write for the ternary compound \mbox{In$_{x}$Ga$_{1-x}$As}
\begin{equation}\label{h-box}
\left\langle h_{j}^{2} \right\rangle = 
4 E_{\mathrm{hf}}^{2} v\int d^{3} R |\psi(\RR)|^{4}
\sum_{i}\frac{I_{i}(I_{i}+1)}{3} q_{i}  \left(\zeta_{i}  \xi_{s,i}^{3}
\mathcal{M}_{j}^{(i)} \right)^{2},
\end{equation}
where $i$ runs through all the nuclear species, 
$q_{i}=(a_{\mathrm{C}}^{(\mathrm{hh})})^{4}r_{i}x$ for In isotopes,
 $q_{i}=(a_{\mathrm{C}}^{(\mathrm{hh})})^{4}r_{i}(1-x)$ for
Ga isotopes and  $q_{i}=(a_{\mathrm{A}}^{(\mathrm{hh})})^{4}$ for
As. The quantity  
\begin{displaymath}
N = \left[v\int d^{3} R |\psi(\RR)|^{4} \right]^{-1}
\end{displaymath}
 is the effective number of the primitive cells encompassed by the wave function
 (the  wave function participation number \cite{Kramer1993}) which
 links the presented theory to 
 the \textit{box model} in which the wave function is considered constant, with the
 value $1/\sqrt{vN}$  over a volume of $N$ unit cells. 

The analogous box-model formula for the electron, which can be inferred directly from
Eq.~\eqref{H6c6c}, is 
\begin{equation}\label{h-box-e}
\left\langle h_{j}^{2} \right\rangle = 
4 E_{\mathrm{hf}}^{2} v\int d^{3} R |\psi(\RR)|^{4}
\sum_{i}\frac{I_{i}(I_{i}+1)}{3} q_{i}  \left(\zeta_{i}  \xi_{s,i}^{3}\right)^{2},
\end{equation}
with $a_{\mathrm{C}}^{(\mathrm{hh})}$ and $a_{\mathrm{A}}^{(\mathrm{hh})}$ in $q_{i}$
replaced by the conduction-band values $a_{\mathrm{C}}^{(\mathrm{e})}$ and
$a_{\mathrm{A}}^{(\mathrm{e})}$, respectively. 

\subsection{Results and discussion}
\label{sec:results}

In this section we study the characteristic strength of the coupling to
longitudinal and transverse fluctuations of the Overhauser field felt
by a hole in the QD (nominally heavy-hole) ground state.
All the results are averages of 10 repetitions in order to account for the random alloying
and isotope distribution, resulting in a standard
deviation of the numerical result on the order of 1\% of the average value.

\begin{figure}[tb]
\begin{center}
\includegraphics[width=\columnwidth]{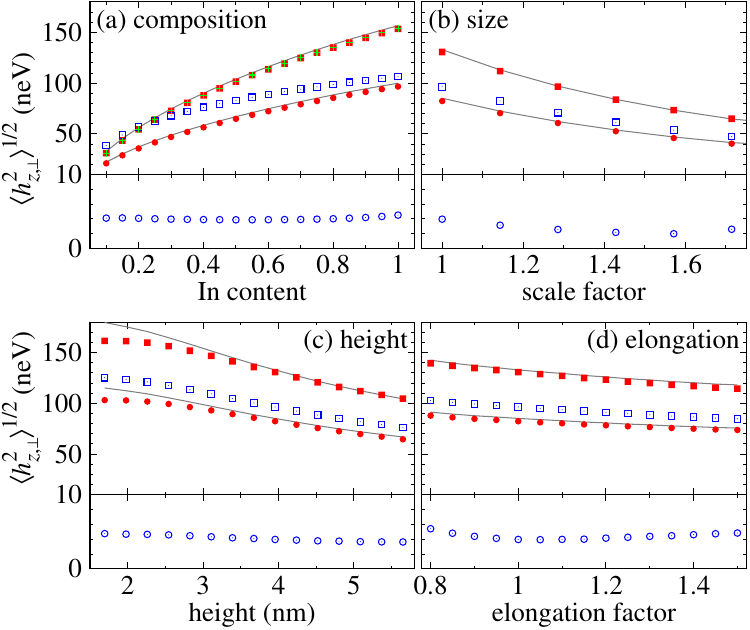}
\end{center}
\caption{\label{fig:results}(Color online) The dependence of the root-mean-square average
  of the   hole hyperfine field fluctuations  for $B=8$~T in the Faraday
  geometry as a function of QD composition (a), size (b), height (c) 
  and shape (d). Squares show the fluctuations of the field component along the growth
  ($z$) axis while circles represent the fluctuations of the transverse components
  (averaged over the in-plane directions). Full red symbols correspond to the model with a
$d$-state admixture to the valence band, while the open blue symbols show the values for
purely $p$-type states. The solid grey lines show the box
model approximation based on the inverse wave function participation number,
given by Eq.~\eqref{h-box}. The green crosses in (a) show the results at
  $B=0.1$~T. The lower part of the vertical axis has been expanded for clarity. }
\end{figure}

Fig.~\ref{fig:results}  shows the results for four series of structures with
different size and composition. The magnetic field is oriented here in the growth
direction (Faraday configuration), hence the $z$ axis is along the symmetry axis of the
structure. 
In our discussion the notions
of `longitudinal' and `transverse' are related to
the growth axis.
Transverse fluctuations are
calculated as the average of fluctuations in two perpendicular directions, 
$\langle h_{\bot}^{2}\rangle^{1/2}=\langle (h_{x}^{2}+h_{y}^{2}))/2\rangle^{1/2}$.

In Fig.~\ref{fig:results}(a) we study cylindrically symmetric lens-shaped QDs
with base radius $21a=11.9$~nm and height $h = 7a=3.96$~nm, and with uniform composition 
\mbox{In$_{x}$Ga$_{1-x}$As}, where the indium content $x$ changes from 0.1 to
1. Without $d$-state admixture to
the valence band and without band-mixing, a heavy hole couples only to longitudinal
hyperfine field. Band mixing induces weak coupling to transverse field (blue open
circles), up to a few percent of the longitudinal one. A much
stronger coupling, comparable to the longitudinal one, appears as a result of $d$-state
admixture (full red circles). The strong dependence on the In content results from the
combination of the large nuclear angular momentum of this element as compared to Ga, and
increasing localization in indium-rich QDs (the  wave function
participation number N decreases 
from $52\cdot 10^{3}$ to $13\cdot 10^{3}$ as $x$ grows from 0.1 to 1). This
dependence is much weaker in the case of transverse coupling induced purely by band
mixing. The grey solid lines show the results obtained from Eq.~\eqref{h-box}. In order to
relate our multi-band numerical wave functions to the simple theory we define here the
 wave function participation number as 
\begin{displaymath}
N' = \left[v\int d^{3} R \left|\sum_{\lambda}|\psi_{\lambda}(\RR)|^{2}\right|^{2} \right]^{-1}
\end{displaymath}
and average the result over the two hh states.
The
agreement is very good, validating the box model with the  wave
  function participation number as the effective number of primitive cells. 
The results for the very weak field of $B=0.1$~T (green crosses) do not differ
  considerably from those at $B=8$~T. This is expected, since in a self-assembled QD the
 in-plane confinement scale ($l_{0}\sim 4$~nm) is much smaller than the magnetic length
 ($l_{B}\approx 9$~nm at $B=8$~T) and the resulting relative field-induced correction to
 confinement ($(l_{0}/l_{B})^{4}/8$ based on the Fock-Darwin model) is negligible. 

Fig.~\ref{fig:results}(b) presents results for a series of QDs with identical
compositions $x=0.75$, starting from the geometry as in the previous case and then
uniformly scaling  each dimension of the QD up by a factor up to 2
(the data is shown as a function of the 
linear scaling factor). In Fig.~\ref{fig:results}(c) the
lateral size of the QD is kept fixed as in Fig.~\ref{fig:results} (a) and the height $h$
is varied. In 
Fig.~\ref{fig:results}(d) the QD is made elliptic by relatively elongating the QD
shape in-plane by a 
fixed factor along the $(1\overline{1}0)$ crystallographic axis while keeping the
  height and the size in the other in-plane direction constant (so that \textit{elongation
    factor} equal to 1 corresponds to the 
  geometry of Fig.~\ref{fig:results}(a)). In all these cases the
general dependence on the geometry qualitatively follows the prediction of the box model
with the effective field fluctuations decreasing with the growing system
size. Quantitatively, however, the fluctuations of the Overhauser field only approximately follow
the expected scaling as $1/\sqrt{V}$, which is due to the fact that the wave function
shrinks slower than the QD when the size of the latter is reduced. In
Fig. ~\ref{fig:results}(c) one can see discrepancy between the numerical values and the
predictions of the box model for very flat QDs. This results from the leakage of the wave
function to the indium-free barrier. 

\begin{figure}[tb]
\begin{center}
\includegraphics[width=\columnwidth]{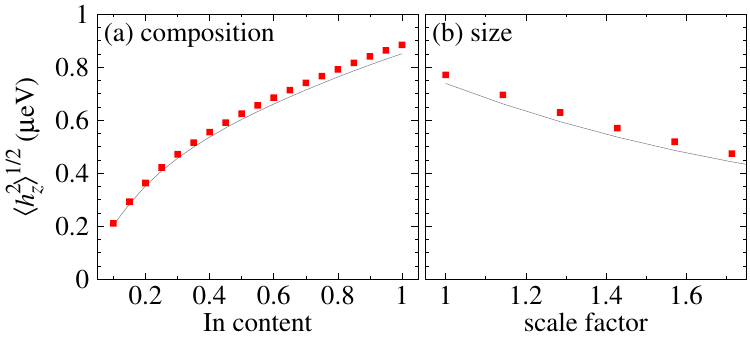}
\end{center}
\caption{\label{fig:results-e}(Color online) The dependence of the root-mean-square average of the
electron hyperfine field fluctuations for $B=8$~T in the Faraday
geometry as a function of QD composition (a) and size (b). Only the field component along the growth
axis is shown. The solid grey lines show the box
model approximation based on the wave function participation number.}
\end{figure}

As a reference, in Fig.~\ref{fig:results-e} we show the rms fluctuations of the Overhauser
field for an electron in the same structures as in Fig.~\ref{fig:results}(a,b). Both the
relative anisotropy of the hyperfine coupling, as well as the relative difference
between the results with and without $d$-shell admixture in this case are at most on the
order of $10^{-3}$, therefore we show only the results for the $z$ component in the model
with the admixture. The values for the electron are 5 to 7 time larger than for the hole,
with the electron-to-hole ratio slightly decreasing as the In content grows. The results for the
electron are also very well reproduced by the box model using the wave function
participation ratio. As the In content grows from $0.1$ to 1, the latter decreases from
$119\cdot 10^{3}$ (more than 
twice the value for the hole in the same structure) to $14\cdot 10^{3}$ (nearly equal to
the hole value).

\begin{figure}[tb]
\begin{center}
\includegraphics[width=0.55\columnwidth]{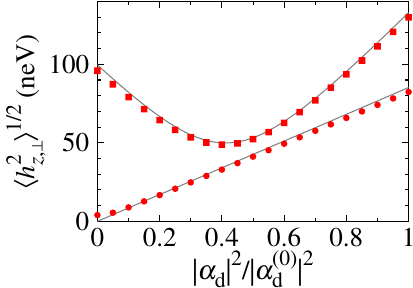}
\end{center}
\caption{\label{fig:d-dependence}(Color online) The dependence of the root-mean-square
  average of the hyperfine field fluctuations on the magnitude of the $d$-shell admixture
  assumed in the calculations. Squares show the fluctuations of the field component along the growth
  ($z$) axis while circles represent the fluctuations of the transverse components
  (averaged over the in-plane directions). The admixtures for all the nuclei are scaled from 0 to the
  values given in Tab.~\ref{params}. The solid grey line shows the box
model approximation based on the wave function participation number,
given by Eq.~\eqref{h-box}. 
%\add{The lower part of the vertical axis has been expanded for clarity}.
} 
\end{figure}

Returning to the holes, one notes that
for the amplitudes of the $d$-state admixtures used here, the magnitudes of the
longitudinal fluctuations with and without the $d$-state admixture are very similar, which
is, however, a coincidence. The dependence of the Overhauser field fluctuations on the
assumed magnitude of $d$-shell admixture is shown in Fig.~\ref{fig:d-dependence}, where we
present the results of calculations with the $d$-shell admixture magnitude for the
nuclear species $i$ set to 
$|\alpha_{d}^{(i)}|^{2} = y |\alpha_{d}^{(i0)}|^{2}$, where $\alpha_{d}^{(i0)}$ are the
values listed in Tab.~\ref{params} and used in the calculations 
presented above, and $0<y<1$. The dependence is non-monotonic. In particular, $y\approx 0.5$
corresponds to mutual compensation of the $p$ and $d$ 
contributions to the coupling to indium ions, which dominate the overall effect due to
their large nuclear momentum. As a result, the longitudinal fluctuations of the effective
field are suppressed.

\begin{figure}[tb]
\begin{center}
\includegraphics[width=\columnwidth]{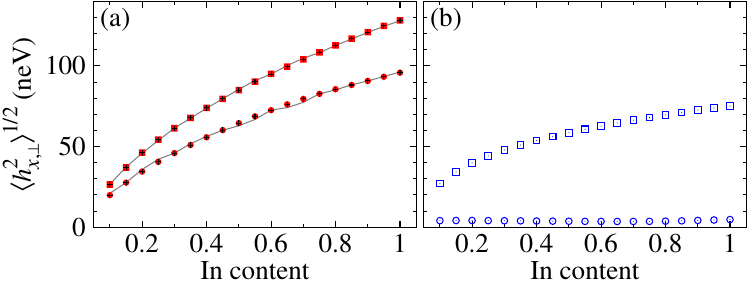}
\end{center}
\caption{\label{fig:Voigt}(Color online) The dependence of the root-mean-square average of the
  hyperfine field fluctuations as a function of QD composition in the Voigt
  geometry with (a) and without (b) atomic $d$-shell admixture.  Circles and squares show the
  fluctuations  along and perpendicular to the magnetic field, respectively, at
$B=8$~T.  Crosses in (a) show the same results at $B=1$~T. For comparison, the grey 
  lines mark the results for the Faraday geometry. }
\end{figure}

As discussed above,
in the strongly confined self-assembled structure the volume occupied by the wave function
depends very weakly on the magnitude and orientation of the magnetic field. Therefore, one
expects that the fluctuations of the Overhauser field will not depend on the
orientation of the magnetic
field.
Fig. 4(a) shows the fluctuations of the Overhauser field in the Voigt geometry (magnetic
field along x). The transverse component of the Overhauser field along the external
magnetic field (the $x$ component, shown by circles) is indeed the same as the transverse
component in the Faraday geometry (shown by a grey line). The fluctuations perpendicular to
the magnetic field (squares) now encompass the longitudinal ($z$) and the other transverse
($y$) component. Again, they perfectly agree with the corresponding average of these two
components in the Faraday geometry (grey Iine).
In addition, we performed computations in the Voigt geometry at $B=1$~T,
shown with crosses in Fig. ~\ref{fig:Voigt}(a). It is clear that the results do not depend
on the field magnitude. 
In Fig. ~\ref{fig:Voigt}(b) we show analogous results from a model assuming no $d$-shell
admixture. As expected, fluctuations perpendicular to the $x$ direction are now much
stronger than the ones along the $x$ axis.

\section{Discussion} \label{sec:discussion}
There two main consequences of our calculations are the following. (1) The effects of
  band mixing on the magnitude of Overhauser field fluctuations experienced by hole spin
  in a self-assembled quantum dot are weak: for most of quantum dot sizes, shapes, and
  compositions one can use a simple single-envelope effective mass wave function to model
  the $z$ component of the Overhauser field. The magnitude of transverse components  of
  the Overhauser field due by band mixing is $<5$ \% of the longitudinal one. (2)
  Inclusion of effects of $d$-state admixture to the hole Bloch functions visibly affects
  the longitudinal fields, and it has an enormous effect on the transverse ones when one
  uses the amplitudes $\alpha_d$ of $d$-state admixtures similar to those inferred in
  \cite{Chekhovich_NP12} from isotope-resolved measurements of the longitudinal Overhauser
  fields caused by dynamically polarized nuclei. Most importantly, for $|\alpha_d|^2$ used
  in \cite{Chekhovich_NP12}, and even for values up to $50$ \% smaller, the Overhauser
  field experienced by the hole spin is almost isotropic.  

Let us discuss the implications of the obtained results for hole spin dephasing in
  Faraday and Voigt configurations. In Faraday configuration, the magnetic field $B$ is
  along the $z$ growth axis of the quantum dot, and the hole spin is initialized in
  superposition of up and down states along the $z$ axis. We assume that the hole spin splitting
  $\Delta E = g_{z}\mu_B B$ (where $g_z$ is the $g$-factor of the hole for $B$ along the
  $z$ axis) is much larger than the transverse Overhauser fields, i.e.~$B \! \gg \! 1$ mT
  assuming $g_{z} \! \approx \! 1$ and $\langle h^{2}_{\perp}\rangle^{1/2} \! \leq \! 100 $~neV. 
% DE = gz * B[mT] * 58 neV
Dephasing of a freely precessing spin is then caused by averaging over a distribution of
Overhauser fields along the $z$ axis (longitudinal field in the terminology of this
paper). The coherence in frame rotating with $\Delta E$ frequency is 
$$ |S_{x}(t) + i S_{y}(t)| \propto \exp[ -(t/\tau_z)^2 ] \,\, $$ 
with $\tau_z \! =\! \sqrt{2}/\sigma_z$, where $\sigma_z \! =\! \langle
h^2_{z}\rangle^{1/2}$. For typical value of $\sigma_z \! \approx \! 100$ neV we have
$T_2^{*} \! \approx \! 9$ ns.  

In the Voigt configuration, with $B$ along the $x$ in-plane direction and $\Delta E
  \! =\! g_{x}\mu_B B \! \gg \! \sigma_z$ (where $g_x$ is the in-plane hole $g$-factor),  
we consider a hole spin initialized in eigenstate of $S_z$ - a superposition of
eigenstates of $\Delta E S_x$. Dephasing of this superposition is caused by averaging over
contributions of $h_x$ to the precession frequency, but also over corrections $(h^2_z +
h^2_y)/2\Delta E$ to this frequency caused by transverse fields
\cite{Fischer_PRB08,Testelin2009a}. In the Faraday configuration such corrections due to
$h^2_\perp/2\Delta E$ were inefficient at dephasing compared to the linear coupling to
$h_z$, since $\langle h^2_\perp \rangle^{1/2} \! < \! \langle h^2_z \rangle^{1/2}$, and
$\Delta E$ is larger by a factor of about $10$ due to anisotropy of hole $g$-factor. In
the Voigt configuration the two mechanisms of dephasing can compete, albeit only at small
magnetic fields. 

Let us first consider the case of almost-isotropic hole hf interaction that we obtain
  using the $d$-state admixture parameters taken from \cite{Chekhovich_NP12}. In this case
  we have $\sigma_z \! \approx \! 150$ neV and $\sigma_{\perp}\! \equiv \! \langle
  h^2_\perp \rangle^{1/2} \approx 100$ neV. Dephasing due to averaging over $h_x$ is
  described by a Gaussian decay with time constant $\tau_{x} \! = \!
  \sqrt{2}/\sigma_{\perp} \! \approx 10$ ns. On the other hand, dephasing due to
  averaging over $h_z$ and $h_y$ fields is described by  
$$ |S_{z}(t) + i S_{y}(t)| \propto \frac{1}{[1+(t/\tau_V)^2]^{1/2}} \,\, , $$ 
with $\tau_V \! \approx \!  \Delta E/\sigma^2_{z}$ (remember that $\sigma_{y}\! \approx \! \sigma_z$ is considered now). With $g_{x} \! \approx \! 0.1$, the
half-decay time following from the above expression is $T_{1/2} \! =\! \sqrt{3}\tau_{V} \!
\approx \!  300$~ns at $B\! =\! 1$ T, and only at fields $<\! 30$ mT this time becomes
shorter than $\tau_x \approx \! 10$ ns, and the coherence is then limited by fluctuations of $h_z$ and
$h_y$. At higher fields the decay is Gaussian with characteristic timescale given by
$\tau_x$.  

On the other hand, in the limit of no $d$-state admixture, we have $\sigma_{\perp} \!
  \approx \! 5$ neV, and the Gaussian decay due to $h_x$ fluctuations occurs in about
  $200$~ns. The mechanism of dephasing due to second-order coupling to $h_z$ (with $h_y$
  fluctuations being now negligible compared to those of $h_z$) leads then  to 
the following form of the coherence decay \cite{Fischer_PRB08,Testelin2009a}
$$ |S_{z}(t) + i S_{y}(t)| \propto \frac{1}{[1+(t/\tau_V)^2]^{1/4}} \,\, , $$ 
which results in half-decay time $T_{1/2} \approx 4\tau_V$. Assuming  $g_{x}\! \approx
0.1$ and $\sigma_z \! \approx \! 150$ neV, for $B\! =\! 1$ T we obtain $T_{1/2}\! \approx
700$~ns, and we see that for $B\! \ll \! 0.3$~T the coherence decay should be dominated
by this mechanism, and it should be possible to observe a characteristic $1/t^2$ tail of
coherence decay when $t\! \ll \! 200$~ns. 
Finally, let us note that in
\cite{Prechtel_NM16} the measured values of rms of Overhauser fluctuations were $\sigma_z
\! \approx \! 60$ neV and $\sigma_\perp \! \approx \! 0.5$ neV and in-plane g-factor of
the hole was $g_{x} \approx 0.05$ (actually $0.035$ for one dot and $0.065$ for
another). These result in coherence half-decay time due to $h_x$ fluctuations given by
$T_{1/2} \approx 2$ $\mu$s at $B\! =\! 1$ T, which means that this mechanism will dominate
over the Gaussian decay at fields already below a Tesla. 

The results that we have obtained using $\alpha_d$ consistent with
  \cite{Chekhovich_NP12} are in clear disagreement with observations presented for InGaAs
  quantum dot in \cite{Prechtel_NM16}, where very small value of transverse Overhauser
  field (smaller by about an order of magnitude than the value that we predict for no $d$ admixture, only
  due to heavy-light hole mixing) was inferred from coherent population trapping
  experiment.   
Our results of small importance of band mixing and good applicability of the box
  model approach to modeling of Overhauser field remove a few possible sources of
  inaccuracies that could have played a role in analysis of measurement results obtained in recent years. This
  strengthens the significance of disagreement in magnitudes of transverse Overhauser
  fields in InGaAs quantum dots inferred from these two very different experiments. 
Analysis in \cite{Chekhovich_NP12} is based on DNP and measurement of isotope-resolved
contributions to the {\it longitudinal} Overhauser shift. The isotope-dependence of signs
of these contributions was explained there by invoking a finite (and in fact quite
substantial) admixture of $d$ orbitals in heavy hole states - but the value of transverse
Overhauser field was not measured in that work. Such a measurement was performed in
\cite{Prechtel_NM16}, in which a possible reason for disagreement with earlier experiments
on dephasing of holes was suggested: the structure used in \cite{Prechtel_NM16} was
carefully designed to exhibit much less charge noise. It is now known that charge noise
can contribute to (or even dominate) hole dephasing dynamics, due to electric-field
dependence of hole $g$-factor \cite{Prechtel_NM16,Huthmacher_PRB18}, so one has to be
careful when attributing observed dephasing to hf interaction and using the measured
coherence time to estimate $\langle h^{2}_{z,\perp}\rangle^{1/2}$. This, however, has no
bearing on the experiment and analysis of \cite{Chekhovich_NP12}.  

\section{Conclusions}
\label{concl}

We have derived the 8-band \kp Hamiltonian for hyperfine interactions, including a
proposed parametrization of Bloch functions consistents with the available experimental data. 
This offers a
general formalism that allows one to include realistic multi-band carrier wave 
functions, as obtained from \kp computations, in the calculation of hyperfine couplings.
Using this formalism, we have studied the effect of fluctuations of the nuclear spin
polarization on a hole in the ground state of an InGaAs QD for a range of realistic shapes and
  sizes, taking into account an admixture of 
atomic $d$ orbitals to the valence band Bloch functions as well as band mixing. 
Our formalism can also be applied to problems in which accurate modeling of carrier
  states is crucial, e.g., when the
hyperfine-related effects are to be combined  with carrier-phonon couplings, compared with
spin-orbit-induced effects or studied in coupled structures where tunneling plays a
role.

One of
  the main results is the observation that in a wide range of dots shapes and sizes, the
  realistic description of carrier states, taking into account band mixing, envelope
  functions leakage into the barrier, etc., has little influence on the root-mean-square of the
  Overhauser field fluctuations experienced by the spin of the heavy hole confined in the
  dot. These fluctuations can be well described using a ``box'' model of wave function,
  with effective number of nuclei strongly coupled to the hole  being the only fitting
  parameter. Such a description was known to hold well for electrons, and it was widely
  used also for holes, but the justification of its quantitative accuracy was lacking
  until now in the latter case. 

For the {\it transverse } (with respect to the growth axis)
fluctuations of the Overhauser field, we have confirmed the relatively small effect of band
mixing as compared to the $d$-state admixture, at least for the magnitude of this
  admixture inferred in \cite{Chekhovich_NP12} from isotope-resolved measurements of
  contributions to the {\it longitudinal} Overhauser field.  
The latter may lead to transverse fluctuations on the same order
of magnitude as the longitudinal ones. 
The dependence of the longitudinal fluctuations on
the amount of $d$ admixture is strong and non-monotonic. In the light of the fact that
  a large variability in the magnitude of both transverse and longitudinal fluctuations
  was reported in experiments, these results suggest the need for careful examination of
  dependence of the magnitude of the $d$-state admixtures to wave functions localized close
  to cation and anion cores (and also the spatial extent of the relevant $d$ orbitals, as it has a large influence on the vlaue of hf interaction), as a function of indium content (and possibly strain) in
  InGaAs/GaAs QDs.  

\acknowledgments
This work was supported by the
Polish National Science Centre under Grant No.~2014/13/B/ST3/04603.
Calculations have been carried out using resources provided by
Wroclaw Centre for Networking and Supercomputing (\url{http://wcss.pl}), Grant No. 203.
We would like to thank Piotr Bogus{\l}awski for interesting discussion of the origins of
$d$-state admixtures in valence band states.

\appendix
\section{Derivation of the matrix elements of the hyperfine coupling}

The main purpose of this Appendix is to rigorously derive the matrix elements of the short-range
multi-band Hyperfine Hamiltonian as given in Eqs.~\eqref{H6c6c}--\eqref{H7v8v}.
In the following, we will focus on one selected nucleus located at
$\bm{R}_{0}$ and the index $\alpha$ will be suppressed. The three
contributions to $\bm{A}$ in Eq.~\eqref{A-all} will be denoted,
respectively, as $\bm{A}_{\mathrm{c}}$ (the contact interaction),
$\bm{A}_{\mathrm{o}}$ (the orbital part of the dipole interaction), and
$\bm{A}_{\mathrm{s}}$ (the spin part of the dipole interaction).

For the general calculations to be performed, it is convenient to use
spherical tensor representation of various vectorial and tensorial
quantities that appear in the derivations. Before we proceed do the technical derivations,
let us note that this is a natural language for discussing the hyperfine spin-flip
selection rules. The essential part of the hyperfine Hamiltonian in
Eq.~\eqref{Hhf} is
\begin{align}\label{AJ}
\bm{A}(\rr-\RR_{0})\cdot\bm{I}&= -\sqrt{3}\sum_{q_{1}q_{2}}
\langle 1,1;q_{1},q_{2} | 1,1;0,0 \rangle \\
& \times A^{(1)}_{q_{1}}(\rr-\RR_{0})I^{(1)}_{q_{2}},\nonumber
\end{align}
where the upper index denotes the rank of the tensor and, at the same time, distinguishes
the spherical tensor components from the cartesian ones,
$\langle j_{1},j_{2};m_{1},m_{2}|j_{1},j_{2};j,m\rangle$ is the
Clebsch-Gordan coefficient and the spherical components of any vector
$\bm{V}$ are defined in the standard way,
\begin{displaymath}
V_{0}^{(1)}=V_{z},\; 
V^{(1)}_{\pm 1} = \frac{\mp  V_{x}-iV_{y}}{\sqrt{2}}
=\mp\frac{1}{\sqrt{2}}V_{\pm}.
\end{displaymath}
The range of $q_{1},q_{2}$ is not given explicitly upon assumption
that ill-defined Clebsch-Gordan coefficients are 0.
Explicitly, 
$\bm{A}\cdot\bm{I} =
A^{(1)}_{0}J^{(1)}_{0}-A^{(1)}_{-1}J^{(1)}_{+1}-A^{(1)}_{+1}J^{(1)}_{-1} =
A_{z}J_{z}+(A_{-}J_{+}+A_{+}J_{-})/2$. The $q=0$ term thus
corresponds to the Ising coupling. The $q=\pm
1$ terms account for spin flip-flop processes, in which the carrier exchanges its spin
with the nucleus.
In the simplest picture of hole states with definite angular momentum and composed
exclusively of $p$ orbitals, the contact part $\bm{A}_{\mathrm{c}}$
does not contribute to valence band hyperfine coupling due to vanishing $p$-type wave
functions at the position of the nucleus. The other two terms can only contribute to
diagonal terms, since the vector operator $\bm{A}$ cannot couple states with
$m_{j}=\pm 3/2$, that is, differing by $|\Delta m_{j}| = 3$. Hence, in this single-band
approximation, only the Ising term appears for heavy holes. However, symmetry
reduction in 
a nanostructure modifies this simple picture by mixing the states belonging to different
representations of angular momentum due to band mixing as well as by admixing $d$-shell
atomic orbitals to valence band Bloch functions.

For the derivations we note that, by comparing Eq.~\eqref{H-kp} with Eq.~\eqref{Hbb}, the
Hamiltonian blocks $\tilde{H}_{b'b}$ contain grouped elements
$3\bm{A}_{\lambda'\lambda}\cdot\bm{I}/(\hbar a^{(\lambda')*}a^{\lambda})$.
We will now derive these elements for each
of the three contributions to the hyperfine Hamiltonian. 

\subsubsection{The contact part}

The contact part, i.e., the first term in Eq.~\eqref{A-all}, has contributions only from
the conduction bands ($s$-type 
atomic orbitals, $l=m=0$). One has
$u_{\mathrm{e}\uparrow}(\rr,\uparrow)=u_{\mathrm{e}\downarrow}(\rr,\downarrow)
=\sqrt{v} a_{\alpha}^{(\mathrm{cb})}  S(r)/\sqrt{4\pi}$, 
$u_{\mathrm{e}\uparrow}(\rr,\downarrow)=u_{\mathrm{e}\downarrow}(\rr,\uparrow)=0$, where
$S_{\alpha}(r)=4\xi_{s}^{3}/a_{\mathrm{B}}^{3}$ is the radial part of the atomic
$s$-type wave function for a given ion. 
Hence, using Eq.~\eqref{All} transformed to spherical tensor components, the contact
interaction has the matrix elements 
\begin{displaymath}
A^{(1)}_{\mathrm{c},q;\lambda'\lambda} = \frac{2\xi_{s}^{3}}{3\hbar}
\left( S_{q}^{(1)} \right)_{s_{\lambda'}s_{\lambda}},
\end{displaymath}
where $s_{\lambda}$ is the spin projection of the electrons in band
$\lambda$. From the Wigner-Eckart theorem one finds
\begin{displaymath}
\left( S_{q}^{(1)} \right)_{s's} = \frac{\sqrt{3}\hbar}{2}
\left\langle\left. \frac{1}{2},1;s,q \right|
  \frac{1}{2},1;\frac{1}{2},s' 
\right\rangle.
\end{displaymath}
Hence, the non-zero matrix elements of the spherical components of the spin
operator are
\begin{align*}
\left(S_{0}^{(1)} \right)_{\uparrow\uparrow} =
-\left(S_{0}^{(1)} \right)_{\downarrow\downarrow} &= \frac{\hbar}{2},\\
-\left(S_{+1}^{(1)} \right)_{\uparrow\downarrow} =
\left(S_{-1}^{(1)} \right)_{\downarrow\uparrow} &= \frac{\hbar}{\sqrt{2}}.
\end{align*}
Collecting the elements of the 6c6c block and converting to Cartesian components one finds 
$A_{\mathrm{c},i}=\xi_{s}^{3} |a_{\alpha}^{(\mathrm{cb})}|^{2} \sigma_{i}/3$, hence 
$\tilde{H}_{6c6c}=3\bm{A}_{c}\cdot\bm{I}/(\hbar |a_{\alpha}^{(\mathrm{cb})}|^{2})=
\bm{\sigma}\cdot\bm{I}/\hbar$, which proves
Eq.~\eqref{H6c6c}. 

\subsubsection{The orbital term of the dipole part}

For the local term of $\bm{A}_{o}$ (the second term in Eq.~\eqref{A-all}) one  
substitutes  the decomposition in Eq.~\eqref{Bloch-i}
into Eq.~\eqref{m-elem-A-sum}. 

The hydrogen-like orbitals $f_{lm}(\rr)$ building the Bloch function according to
Eq.~\eqref{Bloch-i} 
are decomposed into their radial parts $\mathcal{R}(r)$ and
angular parts described by spherical harmonics $Y_{l,m}(\Omega)$.
Taking into account that the components of the angular momentum
operator are diagonal in spin $s$ and in the 
total angular momentum $l$, one gets
\begin{equation*}
\frac{A^{(1)}_{\mathrm{o},q;\lambda'\lambda}}{(a_{\alpha}^{(\lambda')*}a_{\alpha}^{\lambda})}=\xi_{s}^{3}\sum_{lmm's}M_{ll} c^{(\lambda' s)*}_{lm'}c^{(\lambda s)}_{lm}
\langle lm' | L^{(1)}_{q} | lm \rangle,
\end{equation*}
where
\begin{displaymath}
M_{l'l} = \frac{a_{\mathrm{B}}^{3}}{4\xi_{s}^{3}}
\int dr r^{2} \mathcal{R}_{l'}^{*}(\rr) \frac{1}{r^{3}} \mathcal{R}_{l}(\rr).
\end{displaymath}
Following \cite{Chekhovich_NP12} we denote $M_{11}\equiv M_{\mathrm{p}}$
$M_{22}\equiv M_{\mathrm{d}}$,  $M_{02}=M_{20}\equiv  M_{\mathrm{sd}}$.
The matrix elements $\langle lm' | L^{(1)}_{q} | lm \rangle$ can 
be trivially calculated by elementary methods. However, a more compact
and uniform result is obtained via Wigner-Eckart theorem,
\begin{displaymath}
\langle lm' | L^{(1)}_{q} | lm \rangle = 
\frac{\langle l,1;m,q | l,1;l,m' \rangle}{\sqrt{2l+1}}
\langle l || L^{(1)} || l \rangle. 
\end{displaymath}
The reduced matrix element is found by inspection of the component
$q=0$, $m=m'$ where $\langle l,1;m,0 | l,1;lm'
\rangle=m/\sqrt{l(l+1)}$ and obviously $\langle lm | L^{(1)}_{0} | lm
\rangle=m$, hence $\langle l || L^{(1)} || l
\rangle=\sqrt{l(l+1)(2l+1)}$. Hence, the final formula is 
\begin{align}
\lefteqn{\frac{\langle \lambda' | \mathcal{A}^{(1)}_{\mathrm{o,SR},q} |
\lambda \rangle}{(\hbar a_{\alpha}^{(\lambda')*}a_{\alpha}^{\lambda})} =} \label{orbital} \\
&\quad  \xi_{s}^{3}\sum_{lmm's}M_{ll} c^{(\lambda' s)*}_{lm'}c^{(\lambda s)}_{lm}
\sqrt{l(l+1)}\langle l,1;m,q | l,1;l,m' \rangle. \nonumber
\end{align}

We note that this matrix element is diagonal in $l$ and vanishes for $l=0$, hence non-zero
matrix elements appear only within the valence band. 
Moreover, for the heavy-hole (hh) bands,
in the simple single-band approximation, the Bloch functions are spin eigenstates with
opposite spin orientation. Since the orbital contribution is spin-diagonal, in the
single-band, purely $p$-wave model of the hh band, 
this term yields only a diagonal (Ising) coupling.
This coupling  is affected by band mixing only in
the second order, since neither the spin-down nor the $m'=0,-1$ spin-up admixture couple
to the leading-order ($m=1$ spin-up) component of the nominally spin-up hh state
via the $q=0$ tensor component (due to spin conservation and $m+q=m'$ selection rule,
respectively). A $d$-shell admixture introduces a $l=2$, $m'=-1$ spin-up correction to the
spin-up hh state (see the explicit compositions of the Bloch states in
\cite{Chekhovich_NP12}). This is not coupled to the
leading-order ($l=1$) component of this state but couples to the same ($l=2$, $m=-1$)
admixture, leading to a correction to the Overhauser field in the quadratic order in
$\alpha_{d}$. 

In addition, with band mixing, the nominally $+3/2$ (spin-up)
hh state ($m'=1$) may attain an admixture of the spin-down light hole state with
$m'=0$. According to Eq.~\eqref{orbital}, this admixture is coupled to the dominating
component of the $-3/2$ (spin down) hh state ($m=-1$) via the $q=1$ component of
the hyperfine coupling, thus leading to the appearance of spin flip-flop terms in the
Hamiltonian. The $d$-wave admixture to hh Bloch functions are spin-conserving, hence they
can only lead to spin flip-flops in combination with band mixing.

\subsubsection{The spin term of the dipole part}

The third term in Eq.~\eqref{A-all} can be written in terms of cartesian components as 
\begin{displaymath}
A_{\mathrm{s},i}=\sum_{j}T_{ij}S_{j}/\hbar,
\end{displaymath}
where 
\begin{displaymath}
T_{ij}=\frac{a_{\mathrm{B}}^{3}}{4}\frac{3x_{i}x_{j}-r^{2}\delta_{ij}}{r^{5}}
\end{displaymath}
is a traceless, symmetric, second order Cartesian tensor, hence its
components form a second order spherical tensor. 
The spherical components of $\bm{A}_{\mathrm{s}}$ are
\begin{displaymath}
A_{\mathrm{s},q}^{(1)} = -\sqrt{15} \sum_{q_{1},q_{2}}
\langle 2,1;q_{1},q_{2} | 2,1;1,q \rangle
T^{(2)}_{q_{1}}S^{(1)}_{q_{2}}, 
\end{displaymath}
where the spherical components of $T^{(2)}$ are constructed from the
first order position tensor $r^{(1)}$ according to the tensor
multiplication rule,
\begin{align*}
T^{(2)}_{q} & = \frac{a_{\mathrm{B}}^{3}}{4r^{5}} \sum_{q_{1},q_{2}} 
\langle 1,1;q_{1},q_{2} | 1,1;2,q \rangle 
r^{(1)}_{q_{1}} r^{(1)}_{q_{2}} \\
& =\frac{a_{\mathrm{B}}^{3}}{4r^{3}}\sqrt{\frac{8\pi}{15}} Y_{2,q}(\hat{\rr}),
\end{align*}
and the overall factor has been determined by inspection.

The matrix element of $T^{(2)}_{q}$ between two hydrogen-like orbitals is 
\begin{equation}\label{f-T-f}
\int d^{3}r f_{l'm'}^{*}(\rr) T^{(2)}_{q} f_{lm}(\rr) = 
\sqrt{\frac{8\pi}{15}} M_{l'l}   G_{l'2l}^{m'qm},
\end{equation}
where $G_{ll'l''}^{mm'm''}$ are Gaunt coefficients,
\begin{align*}
G_{ll'l''}^{mm'm''} & = \int d\Omega 
Y_{l,m}^{*}(\Omega)Y_{l',m'}(\Omega)Y_{l''m''}(\Omega) \\
& = (-1)^{m}\sqrt{\frac{(2l+1)(2l'+1)(2l''+1)}{4\pi}} \\
&\quad \times
\left(\begin{array}{ccc}
l & l' & l'' \\ 0 & 0 & 0
\end{array}\right)
\left(\begin{array}{ccc}
l & l' & l'' \\ m & m' & m''
\end{array}\right),
\end{align*}
and 
$\left(\begin{array}{ccc}
l & l' & l'' \\ m & m' & m''
\end{array}\right)$
are Wigner 3-$j$ symbols.
From the parity rule on the Gaunt coefficients, $l+l'+l''$ - even, and
the triangle rule, $|l-l''|\le l' \le l+l''$, the only non-zero
contributions in Eq.~\eqref{f-T-f} are those with $(l,l') =
(0,2),(2,0),(1,1),(2,2)$. Hence, non-zero matrix elements appear
within the valence band, and between the valence and conduction
bands. 

Upon substituting the result from Eq.~\eqref{f-T-f}, along with the
decomposition in Eq.~\eqref{Bloch-i}, to Eq.~\eqref{All} one gets
\begin{align}\label{spin}
\lefteqn{\frac{\langle \lambda' | \mathcal{A}^{(1)}_{\mathrm{s,SR},q} |
\lambda \rangle}{(\hbar a_{\alpha}^{(\lambda')*}a_{\alpha}^{\lambda})} 
= -\sqrt{8\pi} \xi_{s}^{3}\sum_{lms}\sum_{l'm's'}M_{l'l}} \\
&\quad \times \sum_{q_{1},q_{2}} \langle 2,1;q_{1},q_{2} | 2,1;1,q \rangle
 G_{l'2l}^{m'q_{1}m}
c^{(\lambda' s')*}_{l'm'}c^{(\lambda s)}_{lm}\left(S^{(1)}_{q_{2}}\right)_{s's}.\nonumber
\end{align}

The structure of this term is much more complicated than that of the orbital contribution, since
the present term is not diagonal in $l$ and $s$. The
Clebsch-gordan coefficient requires $q_{1}+q_{2}=q$, while the
Gaunt and $M_{ll'}$ coefficients impose the selection rules $m'  = m+q_{1}$ and 
$(l',l) = (1,1)$, $(2,2)$, $(0,2)$, or $(2,0)$. 
 
We start with analyzing the corrections to the heavy-hole Overhauser field Ising term,
$q=0$. The three non-vanishing decompositions are now $q_{1}=\pm 1,\,q_{2}=\mp 1$ and
$q_{1}=0,\, q_{2}=0$. 
The contribution $(l',l) = (1,1)$ yields the leading-order (hh-hh) part of the Ising
coupling ($q_{1} = q_{2} = 0$) as well as coupling between components that differ by
spin-orbital angular momentum flip-flop (e.g., $m'=1,\,s'=\uparrow$ to
$m=0,\,s'=\downarrow$). For the $p$-wave component, the latter is only possible for light
hole states, hence the 
resulting correction must rely on  light-hole admixtures to \textit{both} hh states and
is therefore quadratic in band-mixing amplitudes. The contribution $(l',l) = (2,2)$
clearly involves $d$-wave contributions to both states and is therefore always quadratic
in the $d$-wave amplitude $\alpha_{d}$. It contains the contribution from the
leading-order component of the hh state as well as the spin-orbital flip-flop couplings
between light-hole components of the hh state, which are, additionally, quadratic in band mixing. The
term with  $(l',l) = (0,2)$ couples the conduction band (cb) admixture to the $d$-wave
component of the leading-order contribution of the hh state. It is, therefore, linear in
$\alpha_{d}$ but one should remember that the cb admixture is very small. 

In addition, band mixing and $d$-wave contributions to Bloch functions generate terms with
$q=\pm 1$ in the Hamiltonian, that is, flip-flop
couplings between the hh states and the nuclei. For instance, for $(l',l) = (1,1)$,
there is a contribution from the $q_{1}=0$, $q_{2}=1$ term, coupling the leading-order
contribution to the hh state ($m=1$, spin up) with a light-hole admixture to the other hh
state ($m=1$, spin down), which is linear in band mixing and therefore should be much
larger than the band-mixing corrections to the Overhauser term. Another such coupling
appears for $(l',l) = (2,2)$ and $q_{1}=2$, $q_{2}=-1$. This one couples the $m=1$ spin-down
and $m=-1$ spin-up components, that is, the $d$-wave components of the leading
contribution to the two opposite hh states. 

Eq.~\eqref{orbital} together with Eq.~\eqref{spin}, upon converting to Cartesian
components and explicit evaluation, yield the matrix representation used in 
Eqs.~\eqref{H8v8v}--\eqref{H7v8v}.

%\bibliography{refs_decoherence,refs_quant,library}


\end{document}